\documentclass[aps,prl,reprint]{revtex4-1}

\usepackage[utf8x]{inputenc}
\usepackage{graphicx}
\usepackage{hyperref}
\usepackage{amsmath}
\usepackage{float}

\begin{document}

%\preprint{}

\title{Cross-sectional scanning tunneling microscopy of InAs/GaAs(001) submonolayer quantum dots} %Title of paper



\author{R.S.R. Gajjela}
\email[]{r.s.r.gajjela@tue.nl}
%\thanks{}
\affiliation{Department of Applied Physics, Eindhoven University of Technology, Eindhoven 5612 AZ, The Netherlands}
\author{A.L. Hendriks}
\affiliation{Department of Applied Physics, Eindhoven University of Technology, Eindhoven 5612 AZ, The Netherlands}
\author{A.Alzeidan}
\affiliation{Institute of Physics, University of Sao Paulo, Rua do Matao 1371, 05508-090 Sao Paulo, SP, Brazil}
\author{T.F. Cantalice}
\affiliation{Institute of Physics, University of Sao Paulo, Rua do Matao 1371, 05508-090 Sao Paulo, SP, Brazil}
\author{A.A. Quivy}
\affiliation{Institute of Physics, University of Sao Paulo, Rua do Matao 1371, 05508-090 Sao Paulo, SP, Brazil}
\author{P.M. Koenraad}
\affiliation{Department of Applied Physics, Eindhoven University of Technology, Eindhoven 5612 AZ, The Netherlands}
%\email{}

\begin{abstract}
Cross-sectional scanning tunneling microscopy (X-STM) was employed to characterize the InAs submonolayer quantum dots (SMLQDs) grown on top of a Si-doped GaAs(001) substrate in the presence of (2$\times$4) and c(4$\times$4) surface reconstructions. Multiple layers were grown under different conditions to study their effects on the formation, morphology and local composition of the SMLQDs. The morphological and compositional variations in the SMLQDs were observed by both filled and empty-state imaging. A detailed analysis of indium segregation in the SMLQDs layers was described by fitting the local indium-concentration profile with a standard segregation model. We observed a strong influence of the arsenic flux over the indium incorporation and formation of the SMLQDs. We investigated the well-width fluctuations of the InGaAs quantum well (QW) in which the SMLQDs were formed. The well-width fluctuations were small compared to the more pronounced composition fluctuations in all the layers. The lateral compositional variations lead to the formation of indium-rich clusters which can act as quantum dots and yield charge-carrier confinement.
\end{abstract}

%\pacs{}% insert suggested PACS numbers in braces on next line

\maketitle %\maketitle must follow title, authors, abstract and \pacs

\section*{\textbf{Introduction}}
%\label{}
III-V semiconductor quantum dots (QDs) attracted much research interest due to their potential applications in many optoelectronic devices such as lasers \cite{Arakawa1982MultidimensionalCurrent}, single-photon detectors \cite{Fattal2004QuantumSource}, solar cells \cite{Nozik2002QuantumCells}, and quantum information technologies (QITs) \cite{Hadfield2009Single-photonApplications}. Typical III-V self-assembled QDs are grown by molecular beam epitaxy (MBE) in the Stranski-Krastanov (SK) mode where the strain-induced formation of the dots occurs as a consequence of the lattice mismatch between the epitaxial layer and the substrate\cite{Leonard1993DirectSurfaces}. The performance of such SKQDs is limited by their low areal density, high aspect ratio (base to height) and a lower degree of freedom to modify the QD size. Another drawback of SKQDs is the presence of a two-dimensional (2D) wetting layer which reduces the three-dimensional confinement (3D) of the charge carriers.

The submonolayer (SML) technique is considered as an alternative to the SK growth mode to obtain a higher dot density, smaller aspect ratio, to get a better control of their size and composition, and to avoid the formation of the wetting layer\cite{Lenz2011AtomicGaAs,Hopfer2006Single-modeBandwidth}. The main idea behind the growth of SMLQDs is to deposit a fraction of a monolayer (ML) of a low band-gap material forming monolayer-thick islands which are then covered by a few MLs of a high band-gap material and the cycle is repeated. Due to the lattice mismatch between the two materials, a local tensile strain builds up in the thin cap layer which will provide favorable sites to nucleate the islands of the next growth cycle. Under optimal growth conditions, the small islands of multiple layers will stack vertically, behaving like a single quantum dot.  A similar technique has been exploited for the fabrication of columnar quantum dots, starting with a layer of InAs SKQDs, acting  as a seed, followed by cyclic depositions of InAs/GaAs to obtain the desired size of columnar QD\cite{Li2007GrowthSubstrate,He2004FormationGaAs100}.

In general, InAs SMLQDs are grown using experimental conditions very similar to the ones used for InAs SKQDs, i.e, high arsenic flux, low InAs growth rate, and low substrate temperatures.  According to \textit{G.R. Bell et al,}\cite{Bell2000IslandEffects} the nucleation of stable 2D InAs islands occurs only in the presence of a (2$\times$4) reconstruction of the GaAs surface before deposition of the InAs material. However, the combination of a high arsenic flux and low substrate temperature always yields a c(4$\times$4) reconstruction\cite{Kamiya1992Reflectance-differenceVacuum}, where the indium atoms preferentially incorporate into the trenches of the c(4$\times$4) reconstructed surface, randomly alloying the surface rather than forming 2D InAs islands\cite{Belk1997SurfaceEpitaxy}. Therefore, the growth conditions have to be modified, and the arsenic flux has to be considerably reduced to reach the (2$\times$4) surface reconstruction\cite{LaBella2005Arsenic-richStructure}. It has been reported that, due to segregation of indium along the growth direction, indium from the bottom layers segregates and coalesces with indium from the top layers and forms In-rich clusters or agglomerates that behave like quantum dots\cite{Lenz2010AtomicGaAs}. The desired height and composition are achieved by optimizing the number of deposition cycles as well as by optimizing the GaAs-spacer thickness \cite{Lenz2011AtomicGaAs}. The zero-dimensional behavior of SMLQDs has been reported in many works\cite{Harrison2016HeterodimensionalGaAs,Lingnau2016UltrafastDots,Han2019PhotoluminescenceLayers}. Due to their high areal density and better uniformity in size and shape, SMLQDs demonstrated their potential in diverse applications such as 20 Gb/s vertical-cavity surface-emitting lasers (VCSELs) \cite{Hopfer200720Dots,Ledentsov2007SubmonolayerLasers}, diode lasers with high differential gain and lower threshold current density \cite{Mikhrin20000.94Dots}, high detectivity quantum dot infrared photodetectors (QDIPs) \cite{Alzeidan2019High-detectivityReconstruction,Alzeidan2019InvestigationPhotodetector} and solar cells with enhanced open circuit voltage ($V_{OC}$) and reduced short-circuit current density ($J_{SC}$)\cite{Lam2014SubmonolayerCells}.

\section*{Experimental Details}

\begin{figure}
    %\centering
    \includegraphics[width=8.6cm, keepaspectratio]{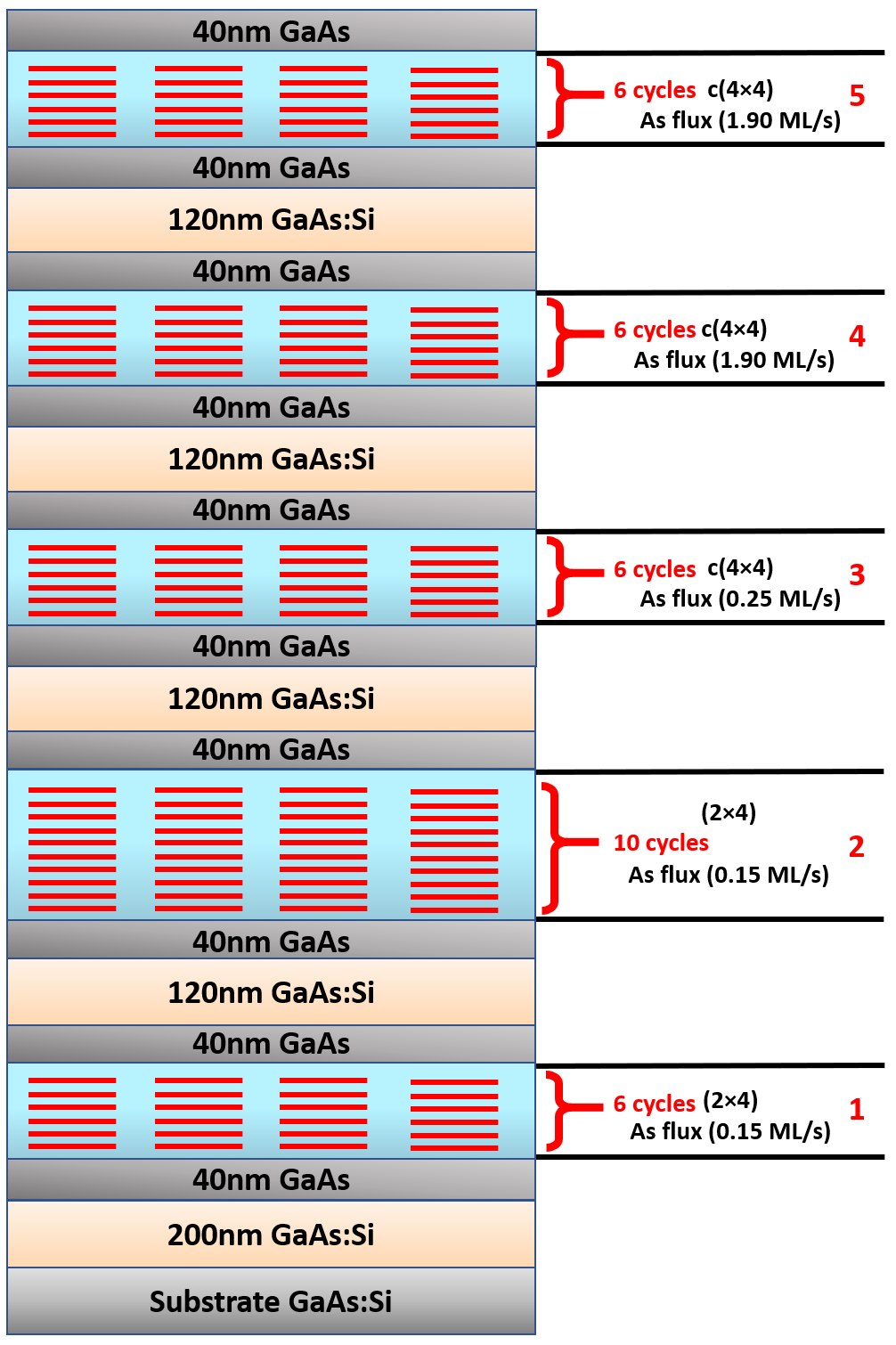}
    \caption{\label{fig:growth}Schematic structure of the X-STM sample grown by MBE}
\end{figure}

The sample was grown by MBE on an epi-ready Si-doped GaAs (001) substrate (doping concentration n=$1×10^{18}$ cm$^{-3}$). After oxide removal and degassing at 600$^o$C during 5 minutes, a 200 nm-thick Si doped GaAs buffer (n=$1×10^{18}$ cm$^{-3}$) was deposited  at 570$^o$C.  In the sequence, five layers containing SMLQDs were grown under different conditions, changing only one parameter at a time (mainly the In, Ga or As flux).  Each SMLQDs layer was deposited at 490$^o$C and consisted of six deposition cycles of 0.5 ML of InAs followed by  2.5 MLs of GaAs (except layer 2 that contained 10 cycles). They were surrounded by 40 nm of GaAs and separated from the next SMLQDs layer by 120 nm of GaAs:Si (n=$1×10^{18}$ cm$^{-3}$) to provide a good conductivity for the X-STM measurements, without having Si atoms inside the nanostructures themselves, as shown in Fig.\ref{fig:growth}.  All these GaAs layers were deposited at 570$^o$C, except for the first 3 nm just above the SMLQDs, that were deposited at 490$^o$C to avoid indium evaporation.  The growth conditions of the five SMLQDs layers were as follows:\\
\noindent \textbf{Layer 1:} The arsenic flux and InAs \& GaAs growth rates were considerably reduced (with respect to the growth conditions normally used for SKQDs) to achieve a (2$\times$4) surface reconstruction. The very low arsenic flux was equivalent to a growth rate of 0.15 ML/s when incident directly onto a Ga-rich surface, while the growth rates of InAs and GaAs were 0.0146 ML/s and 0.1 ML/s respectively. \\ \textbf{Layer 2:} It was grown under the same conditions as layer 1, but ten cycles were deposited instead of six to form the SMLQDs.\\ \textbf{Layer 3:} Same structure and growth conditions as layer 1, except the arsenic flux was slightly increased (0.25 ML/s) in order to obtain a c(4$\times$4) surface reconstruction.\\ \textbf{Layer 4:} The layer was grown same as layer 3, except with a much higher arsenic flux (1.90 ML/s), similar to the one normally used in SKQDs.\\ \textbf{Layer 5:} Same growth conditions as layer 4, but the growth rates of InAs (0.112 ML/s) and GaAs(1.00 ML/s) were increased such that the SMLQDs were grown in the same conditions as usual SKQDs, as often seen in literature.

All the X-STM measurements were performed in a conventional Omicron low-temperature STM at liquid-nitrogen temperature (77K) under ultra-high vacuum ($4-6\times10^{-11}$mbar). The measurements were carried out on a freshly obtained \{110\} surface by cleaving the sample in ultra-high vacuum (UHV). STM tips were made of polycrystalline tungsten wires obtained by electrochemical etching followed by baking and sputtering inside the STM chamber in UHV. All the filled and empty-state images of the SMLQDs layers were acquired in constant current mode. Due to the atomic arrangement of the \{110\} surfaces of Zinc-blende crystals, only every second monolayer along the growth direction is visible in the X-STM images \cite{PhysRevLett.58.1192}. In filled-state imaging at high negative bias voltages, the As sublattice (group V) was imaged, while in empty-state imaging at positive bias voltages, the Ga and In sublattices (group III) were imaged.

\section*{Results and Discussion}

\begin{figure}
    %\centering
    \includegraphics[width=8.6cm, keepaspectratio]{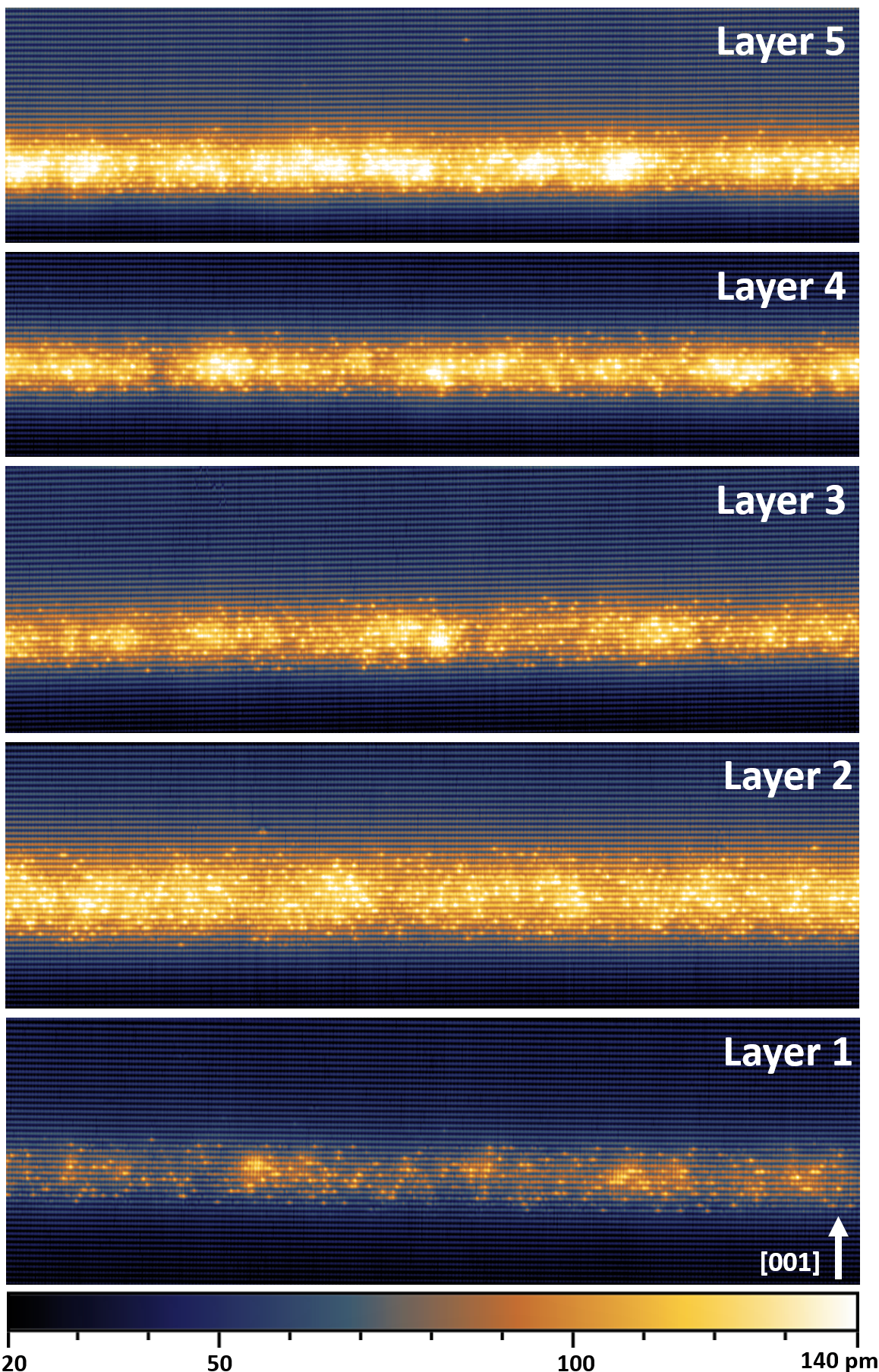}
    \caption{Filled-state images (80$\times$25 nm$^2$) of all the SMLQDs layers 1 to 5 taken at bias voltage $(V_b)$ = - 2.1 V and tunneling current $(I_t)$ = 50 pA. The arrow indicates the growth direction [001].}
    \label{fig:filled}
\end{figure}

\begin{figure}
    %\centering
    \includegraphics[width=8.6cm, keepaspectratio]{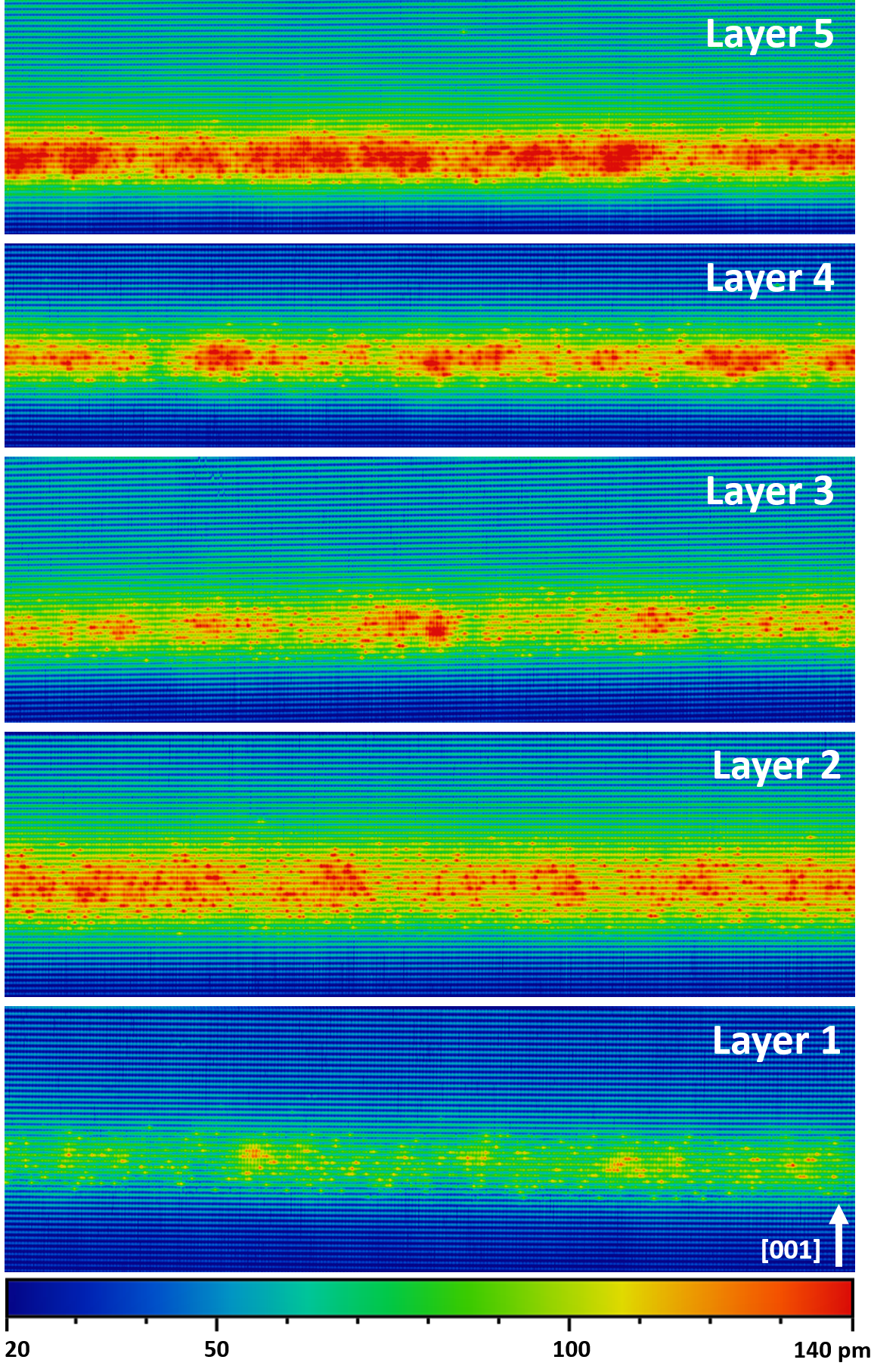}
    \caption{Same filled-state images (80$\times$25 nm$^2$) as in Fig.\ref{fig:filled}, but with a different contrast, taken at $V_b$ = - 2.1 V and $I_t$ = 50 pA, showing the In-rich clusters (bright red regions).}
    \label{fig:contrast}
\end{figure}

Fig.\ref{fig:filled} shows atomically resolved topographic filled-state X-STM images of all five SMLQDs layers obtained at high negative bias voltage. From the morphological point of view, there is a clear difference in the appearance of bright regions among the layers, the bright contrast in the image being due to the outward relaxation of the compressively strained InAs regions after cleavage. From the filled-state images, there is no clear separation of the InAs submonolayer (0.5 ML) and GaAs spacer (2.5 MLs), indicating the segregation of indium atoms over several monolayers along the growth direction [001].
From Fig.\ref{fig:filled}, there is a clear difference in the brightness along the lateral direction, indicating inhomogeneity in the indium composition within the layers, and suggesting the formation of well-separated In-rich clusters. The color contrast of the filled-state images was modified in Fig.\ref{fig:contrast} to better show the compositional fluctuations and indium clustering.

The lateral compositional fluctuations are more pronounced in layer 1 (grown with a (2$\times$4) surface reconstruction) than in the other layers grown with a c(4$\times$4) reconstruction, although there is no clear indication of vertical stacking of 2D InAs islands for the formation of SMLQDs (see Fig.\ref{fig:contrast}). Layer 2, which was grown in the same way as layer 1, but contains 10 cycles instead of 6, looks similar to layer 1 without any indication of SMLQDs formation. Despite the fact that they have the same indium concentration, layer 2 appears brighter than layer 1 due to the enhanced surface relaxation that increases with the total thickness of the InGaAs layer.  Layers 3, 4, and 5, which were grown with a c(4$\times$4) surface reconstruction due to the higher arsenic flux, show a systematic tendency to form indium-rich clusters. Among all the SMLQDs layers, layer 5, grown at high arsenic, gallium and indium fluxes, shows the highest density of 5-6$\times$10$^{11}$ SMLQDs/cm$^2$, which is roughly 10 times the usual SKQDs density. The SMLQDs have a base length of 4\textendash6 nm and a height of 3\textendash3.5 nm, which lead to a smaller aspect ratio than for SKQDs that contributes to enhance the efficiency of the devices\cite{Kim2015Multi-stackedCells}. In the filled-state images (Fig.\ref{fig:filled} and Fig.\ref{fig:contrast}), the intensity of the bright regions increases from layer 1 to 5, indicating a variation in indium concentration among the layers, although all of them nominally received exactly the same amount of indium. In general, the growth temperature has a significant effect on the incorporation of indium atoms into the bulk\cite{Muraki1992SurfaceWells}.  However, in our case, the five SMLQDs layers were deposited at the same temperature (490$^o$C), and the substrate temperature cannot be responsible for this behavior. From the X-STM analysis, it seems that the arsenic flux is also an important parameter, as well as the GaAs and InAs growth rates.  A common point to all these observations is the indium segregation, which is a physical phenomenon that limits indium incorporation, is a function of the growth parameters and is weaker when the arsenic flux and the growth rates increase\cite{Muraki1992SurfaceWells, Martini2003InfluenceSurface}.

\begin{figure}
    %\centering
    \includegraphics[width=8.6cm, keepaspectratio]{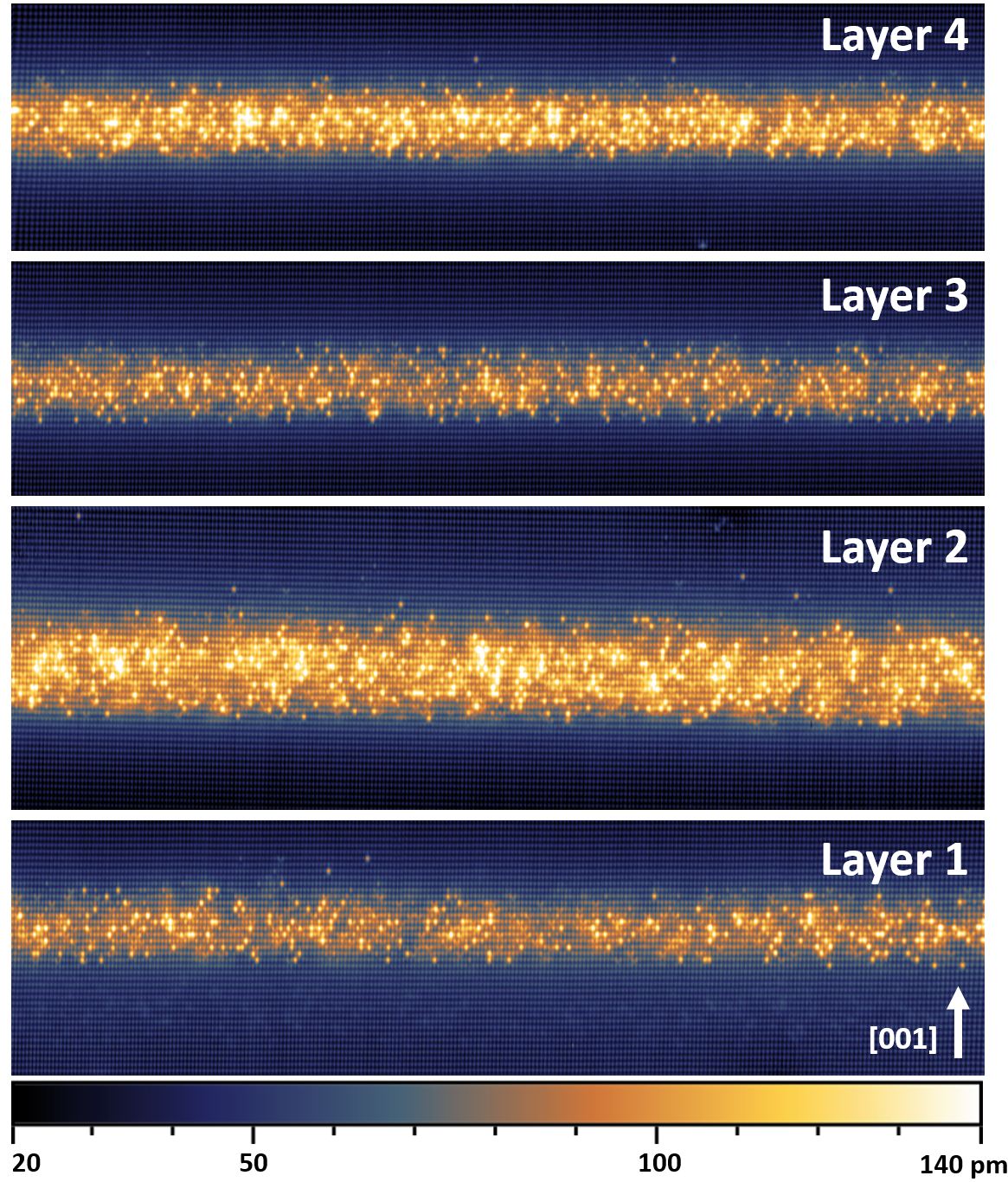}
    \caption{Atomic-resolution empty-state images (80$\times$25 nm$^2$) of the SMLQDs layers 1 to 4, taken at $V_b$ = + 2.5 V and $I_t$ = 50 pA. The arrow indicates the growth direction [001].}
    \label{fig:empty}
\end{figure}

\begin{figure}
    %\centering
    \includegraphics[width=8.6cm, keepaspectratio]{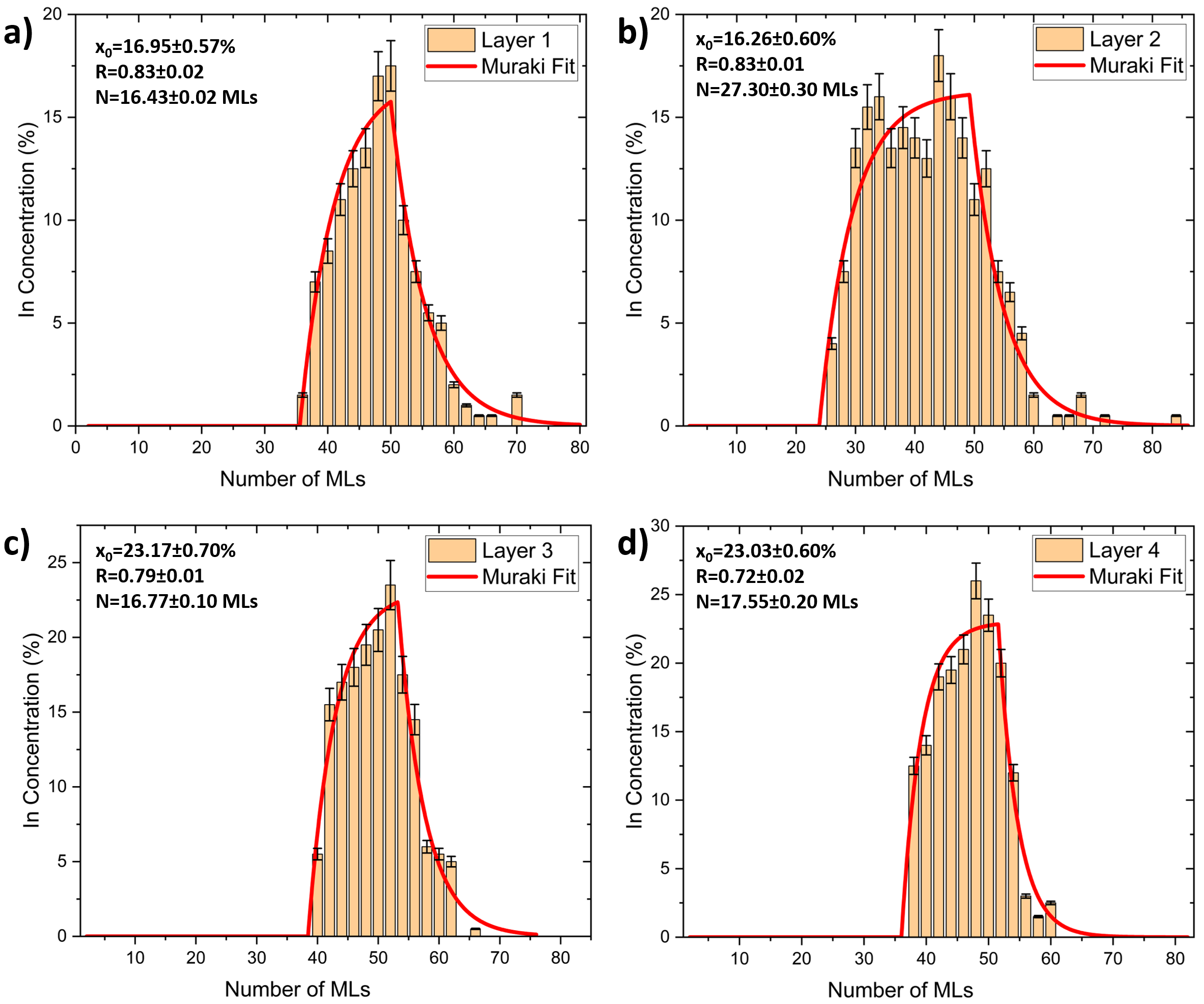}
    \caption{Indium-concentration profile of the layers of Fig.\ref{fig:empty} along the growth direction [001]: (a) Layer 1, (b) Layer 2, (c) Layer 3, (d) Layer 4. The best fits of the data using the phenomenological segregation model of Muraki are also shown (red) where R is the segregation coefficient or probability, N is the number of monolayers containing indium atoms that were deposited (well width) and $x_o$ is the nominal indium concentration.}
    \label{fig:muraki}
\end{figure}

To further study the effect of the arsenic flux on the local composition of the SMLQDs, empty-state imaging was performed. Fig.\ref{fig:empty} shows atomically resolved empty-state images of the SMLQDs layers 1 to 4, at positive bias voltage revealing the group III sublattice. Due to the tip drop off over the edge during the measurement, it was impossible to get the empty-state images of layer 5.
The individual indium atoms are identified as bright point-like features in the images. The local indium concentration of the SMLQDs layers was estimated on each atomic corrugation line.
The extent of indium segregation and nominal indium concentration in each layer were obtained by fitting the indium concentration profile with the phenomenological model of \textit{Muraki et al.,}\cite{Muraki1992SurfaceWells} which allowed to derive the segregation coefficient R (the fraction of indium atoms that segregate from one monolayer to the next one):

\begin{equation*}
  x_{n} =
    \begin{cases}
      0 \quad for \quad n < 1 \\
      x_0(1-R^n) \quad for  \quad 1\leq n \leq N \\
      x_0(1-R^N) R^{n-N}, \quad for \quad n > N
    \end{cases}       
\end{equation*}
In this model, $x_n$ is the actual indium concentration inside the n$^{th}$ monolayer, N is the total number of monolayers containing indium atoms that were deposited (well width) and $x_0$ is the nominal indium concentration. Fig.\ref{fig:muraki} shows the number of indium atoms per atomic layer along the growth direction and a fit to the data (using above equations) from which the segregation coefficient (R), well width (N) and nominal indium concentration ($x_0$) were determined.

As expected from the filled-state images, there is a strong segregation of indium along the growth direction at lower arsenic flux. From the fitting, one can see that layers 1 and 2 have the same segregation coefficient (R=0.83), which is consistent with the fact that they were grown under the same conditions, the only difference being the number of cycles (6 and 10, respectively). When the arsenic flux is increased, as in layers 3 and 4, the segregation coefficient was reduced to 0.79 and 0.72.  Although it was not possible to get empty-state images  of layer 5, we expect that its segregation coefficient would be even lower (as suggested in Fig.\ref{fig:filled} and Fig.\ref{fig:contrast}), as the InAs and GaAs growth rates were much higher than in the other layers, and also contribute to the reduction of indium segregation. Such values of R are in  excellent agreement with the ones obtained independently, by in-situ reflection high-energy electron diffraction (RHEED), during the growth of samples under conditions similar to the ones used here\cite{Cantalice2019In-situDots}. The well width (N) derived from the fitting is in rather good agreement with the nominal number of atomic layers deposited to form the  SMLQDs (18 MLs for layers 1,3, \& 4, and 30 MLs for layer 2). Another interesting result from the fitting is that the indium concentration ($x_0$) of the InAs layers that were deposited for the formation of the SMLQDs is lower in layers 1 and 2 ($x_0$=0.16) than in layers 3 and 4 (0.23).  Since the deposition cycles were exactly the same for all the layers, one should expect $x_0$ to be constant for all of them, irrespective of the  segregation coefficient. The indium atoms that are not incorporated at once, will segregate and be incorporated in the next layers, as can be seen in  the equations above.  Therefore, this variation of $x_0$ is not related to any segregation effect but rather to a physical limitation of the incorporation process of the indium atoms themselves.  Since the (2$\times$4) surface  reconstruction could only be achieved with a very low arsenic flux, we believe that such a flux is also responsible for the reduction of indium incorporation\cite{Muraki1992SurfaceWells}, leading later to the evaporation of indium atoms from the surface.  This is why layers 1 and 2 contain less indium atoms than the other layers and look less bright.

\begin{figure*}
    %\centering
    \includegraphics[width=\textwidth, height=8cm, keepaspectratio]{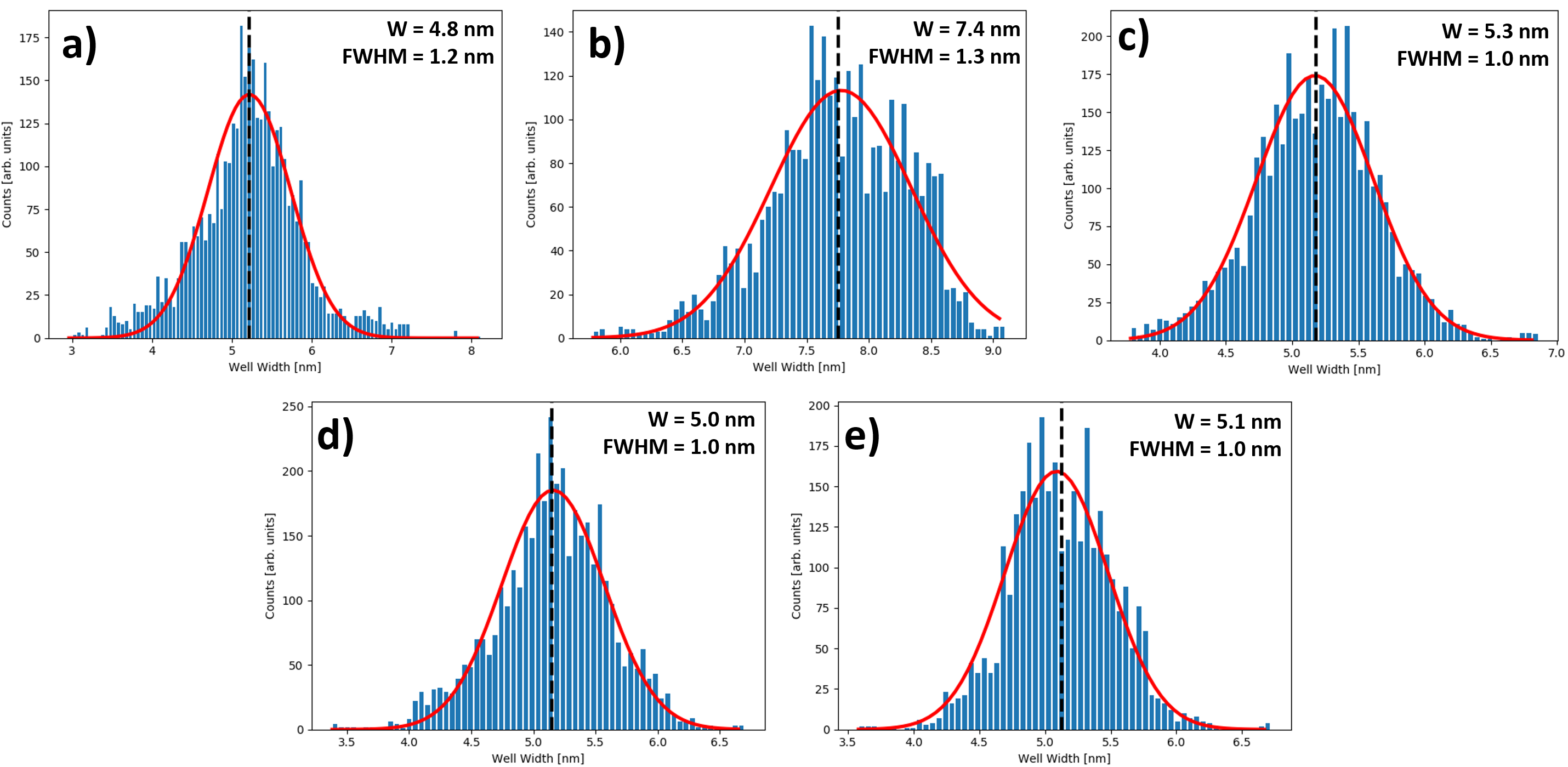}
    \caption{Distribution of well widths summed  over a distance of 400 nm for each SMLQD layer with a Gaussian fit(red).  The average well width (W) and full-width at half maximum (FWHM) are indicated: (a) Layer 1; (b) Layer 2; (c) layer 3; (d) Layer 4; (e) Layer 5.}
    \label{fig:fwhm}
\end{figure*}

It is clear from the X-STM images that the InAs/GaAs SMLQDs can not be considered as vertical stacks of 2D InAs islands surrounded by pure GaAs material.  Instead, they look like In-rich clusters that are embedded into a thick InGaAs quantum well (QW) having a lower indium content. It is also clear that the SMLQDs don't have a more homogeneous size distribution than SKQDs, as often believed as a result of their narrower photoluminescence spectrum\cite{Harrison2016HeterodimensionalGaAs}. The roughness at both interfaces of this quantum well introduces local fluctuations of the potential well that hinder the motion of charge carriers and affect the spectral line width and the recombination rate of the devices\cite{Uzdavinys2017InfluencePhotoluminescence}.  
\begin{figure}
    \includegraphics[width=8.6cm, keepaspectratio]{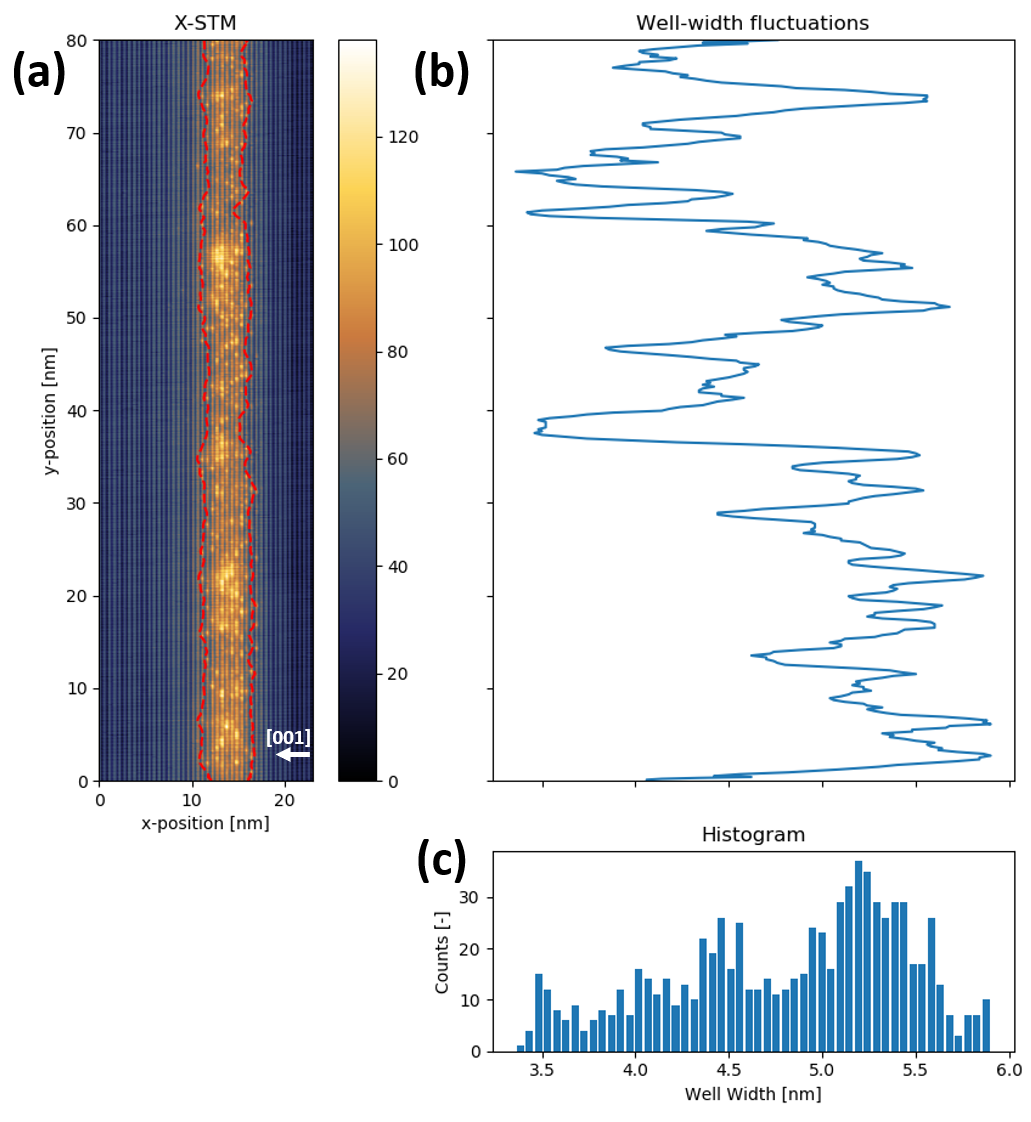}
    \caption{Well-width analysis: (a) X-STM filled-state image of layer 1 showing the interface on both sides of the InGaAs QW (red lines); (b) local well-width fluctuations of the QW corresponding to the horizontal distance between the two red lines; (c) histogram displaying the distribution of well widths across the whole image.}
    \label{fig:well}
\end{figure} 
The monolayer fluctuations in the InGaAs-QW width were observed by measuring the thickness at each pixel row perpendicular to the QW in the filled-state images of all SMLQD layers. Fig.\ref{fig:well} shows an example of the fluctuations of the InGaAs-well width over a length of 80 nm in layer 1 (see Fig.S.1 to Fig.S.4 in Supplemental Material for layers 2 to 5). The well width fluctuations were summed over a distance of 400 nm along the lateral direction for all the five layers using multiple filled-state images taken at the same tunneling conditions. The distribution of the well widths shows a near Gaussian behavior for all the SMLQDs, as illustrated in Fig.\ref{fig:fwhm}, where it can also be seen that they all have an average full-width at half maximum (FWHM) of around 1 nm. 
The average well width of each distribution is in good agreement with the well-width values (N) calculated from the fits presented in Fig.\ref{fig:muraki}. The similarity between FWHM results indicates that the well-width fluctuations do not depend on the growth conditions. Since their values are quite small, they will have a very weak effect on charge-carrier confinement. From the X-STM analysis, it is very clear that the lateral composition fluctuations lead to the formation of In-rich clusters, will have a major contribution to the 3D confinement of charge carriers.

Despite the expectation that the (2$\times$4) reconstruction (layers 1 \& 2) should allow the formation of 2D InAs islands\cite{Bell2000IslandEffects}, the X-STM results show no such islands formation under the current growth conditions. The incorporation of indium is lower than expected in the presence of a (2$\times$4) surface reconstruction due to the very low arsenic flux. On the other hand, layers (3, 4, \& 5) deposited with a c(4$\times$4) surface construction using a higher arsenic flux yielded a higher density of indium-rich clusters, probably due to the lower segregation coefficient and higher indium incorporation rate. However, one can observe that the SMLQDs don't have the full expected height (18 MLs for layers 1, 3, 4 and 5).  That can be explained by the strain in the SMLQDs that seems to be too small to induce vertical stacking, when compared to typical SKQDs\cite{Li2007GrowthSubstrate}, due to the lower amount of indium in the layers. The local indium-rich clusters with a base length of 4\textendash6 nm and a height of 3\textendash3.5 nm can lead to charge-carrier confinement similar to the one of SKQDs of comparable dimensions\cite{Pohl2005EvolutionDots}. Although  SMLQDs are supposed to have no wetting layer, the InGaAs QW in which they are embedded could actually act as one, since it is a 2D system directly in contact with the SMLQDs. Hetero-dimensional confinement was reported in SMLQDs, with the electron wavefunction extending over multiple indium-rich clusters in the InGaAs QW, while the hole was 3D confined inside a single SMLQD\cite{Harrison2016HeterodimensionalGaAs}.
Even though SMLQDs layers grown with a c(4$\times$4) reconstruction at higher arsenic flux show a higher density of nanostructures, the  growth conditions to get SMLQDs in the presence of a (2$\times$4) surface reconstruction might still be optimized to increase In incorporation and decrease segregation. Perhaps then it would be possible to observe the nucleation of 2D InAs islands\cite{He2004FormationGaAs100} which might vertically stack to form SMLQDs having better structural and optoelectronic properties. 

\section*{Conclusions}
In conclusion, we investigated multiple layers of InAs/GaAs submonolayer quantum dots grown on a Si-doped GaAs(001) substrate under different conditions by X-STM with atomic resolution. The morphology and local indium composition were studied by filled-state and empty-state imaging. The arsenic flux had a strong impact on the formation of SMLQDs than surface reconstruction. From the X-STM images, we could not observe any nucleation of 2D InAs islands in the presence of a (2$\times$4) surface reconstruction. The SMLQDs layers grown at high arsenic flux, higher InAs and GaAs growth rates with a c(4$\times$4) surface reconstruction formed highly dense (5-6$\times$10$^{11}$ SMLQDs/cm$^2$) In-rich clusters that can mimic quantum dots. Increasing the arsenic flux significantly decreases indium segregation during the growth. The interface roughness of the InGaAs well width is not significant compared to the more pronounced compositional fluctuations. Thus, the present work provides a detailed insight into the fundamental understanding of growth and formation of SMLQDs, that is needed for further growth optimization.
\section*{acknowledgments}
The research was supported by the funding from the European Union’s Horizon 2020 research and innovation program under the Marie Skłodowska-Curie, project 4PHOTON grant agreement No 721394 and by the Coordenação de Aperfeiçoamento de Pessoal de Nível Superior - Brasil (CAPES) - Finance Code 001, and CNPq (grant 311687/2017-2). The authors thank Alexey Federov and Artur Tuktamyshev of L-NESS laboratory, UNIMI-Bicocca, Italy, for the fruitful discussions about the growth of submonolayer quantum dots.

% If in two-column mode, this environment will change to single-column format so that long equations can be displayed. 
%\begin{widetext}
%$$\mbox{put long equation here}$$
%\end{widetext}

% Figures should be put into the text as floats. 
% Use the graphics or graphicx packages (distributed with LaTeX2e).
% Put the label that you will use with \ref{} command in the braces of the \label{} command.
%
% \begin{figure}
% \end{figure}

% \begin{table}
% \end{tabular}
% \end{table}

%\begin{acknowledgments}
% Put your acknowledgments here.
%\end{acknowledgments}

\bibliographystyle{apsrev4-1}
\bibliography{references}

\begin{thebibliography}{30}%
\makeatletter
\providecommand \@ifxundefined [1]{%
 \@ifx{#1\undefined}
}%
\providecommand \@ifnum [1]{%
 \ifnum #1\expandafter \@firstoftwo
 \else \expandafter \@secondoftwo
 \fi
}%
\providecommand \@ifx [1]{%
 \ifx #1\expandafter \@firstoftwo
 \else \expandafter \@secondoftwo
 \fi
}%
\providecommand \natexlab [1]{#1}%
\providecommand \enquote  [1]{``#1''}%
\providecommand \bibnamefont  [1]{#1}%
\providecommand \bibfnamefont [1]{#1}%
\providecommand \citenamefont [1]{#1}%
\providecommand \href@noop [0]{\@secondoftwo}%
\providecommand \href [0]{\begingroup \@sanitize@url \@href}%
\providecommand \@href[1]{\@@startlink{#1}\@@href}%
\providecommand \@@href[1]{\endgroup#1\@@endlink}%
\providecommand \@sanitize@url [0]{\catcode `\\12\catcode `\$12\catcode
  `\&12\catcode `\#12\catcode `\^12\catcode `\_12\catcode `\%12\relax}%
\providecommand \@@startlink[1]{}%
\providecommand \@@endlink[0]{}%
\providecommand \url  [0]{\begingroup\@sanitize@url \@url }%
\providecommand \@url [1]{\endgroup\@href {#1}{\urlprefix }}%
\providecommand \urlprefix  [0]{URL }%
\providecommand \Eprint [0]{\href }%
\providecommand \doibase [0]{http://dx.doi.org/}%
\providecommand \selectlanguage [0]{\@gobble}%
\providecommand \bibinfo  [0]{\@secondoftwo}%
\providecommand \bibfield  [0]{\@secondoftwo}%
\providecommand \translation [1]{[#1]}%
\providecommand \BibitemOpen [0]{}%
\providecommand \bibitemStop [0]{}%
\providecommand \bibitemNoStop [0]{.\EOS\space}%
\providecommand \EOS [0]{\spacefactor3000\relax}%
\providecommand \BibitemShut  [1]{\csname bibitem#1\endcsname}%
\let\auto@bib@innerbib\@empty
%</preamble>
\bibitem [{\citenamefont {Arakawa}\ and\ \citenamefont
  {Sakaki}(1982)}]{Arakawa1982MultidimensionalCurrent}%
  \BibitemOpen
  \bibfield  {author} {\bibinfo {author} {\bibfnamefont {Y.}~\bibnamefont
  {Arakawa}}\ and\ \bibinfo {author} {\bibfnamefont {H.}~\bibnamefont
  {Sakaki}},\ }\href {\doibase 10.1063/1.92959} {\bibfield  {journal} {\bibinfo
   {journal} {Appl. Phys. Lett}\ }\textbf {\bibinfo {volume} {40}},\ \bibinfo
  {pages} {939} (\bibinfo {year} {1982})}\BibitemShut {NoStop}%
\bibitem [{\citenamefont {Fattal}\ \emph {et~al.}(2004)\citenamefont {Fattal},
  \citenamefont {Diamanti}, \citenamefont {Inoue},\ and\ \citenamefont
  {Yamamoto}}]{Fattal2004QuantumSource}%
  \BibitemOpen
  \bibfield  {author} {\bibinfo {author} {\bibfnamefont {D.}~\bibnamefont
  {Fattal}}, \bibinfo {author} {\bibfnamefont {E.}~\bibnamefont {Diamanti}},
  \bibinfo {author} {\bibfnamefont {K.}~\bibnamefont {Inoue}}, \ and\ \bibinfo
  {author} {\bibfnamefont {Y.}~\bibnamefont {Yamamoto}},\ }\href {\doibase
  10.1103/PhysRevLett.92.037904} {\bibfield  {journal} {\bibinfo  {journal}
  {Physical Review Letters}\ }\textbf {\bibinfo {volume} {92}},\ \bibinfo
  {pages} {4} (\bibinfo {year} {2004})}\BibitemShut {NoStop}%
\bibitem [{\citenamefont {Nozik}(2002)}]{Nozik2002QuantumCells}%
  \BibitemOpen
  \bibfield  {author} {\bibinfo {author} {\bibfnamefont {A.~J.}\ \bibnamefont
  {Nozik}},\ }in\ \href {\doibase 10.1016/S1386-9477(02)00374-0} {\emph
  {\bibinfo {booktitle} {Physica E: Low-Dimensional Systems and
  Nanostructures}}},\ Vol.~\bibinfo {volume} {14}\ (\bibinfo {year} {2002})\
  pp.\ \bibinfo {pages} {115--120}\BibitemShut {NoStop}%
\bibitem [{\citenamefont
  {Hadfield}(2009)}]{Hadfield2009Single-photonApplications}%
  \BibitemOpen
  \bibfield  {author} {\bibinfo {author} {\bibfnamefont {R.~H.}\ \bibnamefont
  {Hadfield}},\ }\href {\doibase 10.1038/nphoton.2009.230} {\enquote {\bibinfo
  {title} {{Single-photon detectors for optical quantum information
  applications}},}\ } (\bibinfo {year} {2009})\BibitemShut {NoStop}%
\bibitem [{\citenamefont {Leonard}\ \emph {et~al.}(1993)\citenamefont
  {Leonard}, \citenamefont {Krishnamurthy}, \citenamefont {Reaves},
  \citenamefont {Denbaars},\ and\ \citenamefont
  {Petroff}}]{Leonard1993DirectSurfaces}%
  \BibitemOpen
  \bibfield  {author} {\bibinfo {author} {\bibfnamefont {D.}~\bibnamefont
  {Leonard}}, \bibinfo {author} {\bibfnamefont {M.}~\bibnamefont
  {Krishnamurthy}}, \bibinfo {author} {\bibfnamefont {C.~M.}\ \bibnamefont
  {Reaves}}, \bibinfo {author} {\bibfnamefont {S.~P.}\ \bibnamefont
  {Denbaars}}, \ and\ \bibinfo {author} {\bibfnamefont {P.~M.}\ \bibnamefont
  {Petroff}},\ }\href {\doibase 10.1063/1.110199} {\bibfield  {journal}
  {\bibinfo  {journal} {Applied Physics Letters}\ }\textbf {\bibinfo {volume}
  {63}},\ \bibinfo {pages} {3203} (\bibinfo {year} {1993})}\BibitemShut
  {NoStop}%
\bibitem [{\citenamefont {Lenz}\ \emph {et~al.}(2011)\citenamefont {Lenz},
  \citenamefont {Eisele}, \citenamefont {Becker}, \citenamefont {Schulze},
  \citenamefont {Germann}, \citenamefont {Luckert}, \citenamefont
  {P{\"{o}}tschke}, \citenamefont {Lenz}, \citenamefont {Ivanova},
  \citenamefont {Strittmatter}, \citenamefont {Bimberg}, \citenamefont {Pohl},\
  and\ \citenamefont {D{\"{a}}hne}}]{Lenz2011AtomicGaAs}%
  \BibitemOpen
  \bibfield  {author} {\bibinfo {author} {\bibfnamefont {A.}~\bibnamefont
  {Lenz}}, \bibinfo {author} {\bibfnamefont {H.}~\bibnamefont {Eisele}},
  \bibinfo {author} {\bibfnamefont {J.}~\bibnamefont {Becker}}, \bibinfo
  {author} {\bibfnamefont {J.-H.}\ \bibnamefont {Schulze}}, \bibinfo {author}
  {\bibfnamefont {T.~D.}\ \bibnamefont {Germann}}, \bibinfo {author}
  {\bibfnamefont {F.}~\bibnamefont {Luckert}}, \bibinfo {author} {\bibfnamefont
  {K.}~\bibnamefont {P{\"{o}}tschke}}, \bibinfo {author} {\bibfnamefont
  {E.}~\bibnamefont {Lenz}}, \bibinfo {author} {\bibfnamefont {L.}~\bibnamefont
  {Ivanova}}, \bibinfo {author} {\bibfnamefont {A.}~\bibnamefont
  {Strittmatter}}, \bibinfo {author} {\bibfnamefont {D.}~\bibnamefont
  {Bimberg}}, \bibinfo {author} {\bibfnamefont {U.~W.}\ \bibnamefont {Pohl}}, \
  and\ \bibinfo {author} {\bibfnamefont {M.}~\bibnamefont {D{\"{a}}hne}},\
  }\href {\doibase 10.1116/1.3602470} {\bibfield  {journal} {\bibinfo
  {journal} {Journal of Vacuum Science {\&} Technology B, Nanotechnology and
  Microelectronics: Materials, Processing, Measurement, and Phenomena}\
  }\textbf {\bibinfo {volume} {29}},\ \bibinfo {pages} {04D104} (\bibinfo
  {year} {2011})}\BibitemShut {NoStop}%
\bibitem [{\citenamefont {Hopfer}\ \emph {et~al.}(2006)\citenamefont {Hopfer},
  \citenamefont {Mutig}, \citenamefont {Kuntz}, \citenamefont {Fiol},
  \citenamefont {Bimberg}, \citenamefont {Ledentsov}, \citenamefont {Shchukin},
  \citenamefont {Mikhrin}, \citenamefont {Livshits}, \citenamefont
  {Krestnikov}, \citenamefont {Kovsh}, \citenamefont {Zakharov},\ and\
  \citenamefont {Werner}}]{Hopfer2006Single-modeBandwidth}%
  \BibitemOpen
  \bibfield  {author} {\bibinfo {author} {\bibfnamefont {F.}~\bibnamefont
  {Hopfer}}, \bibinfo {author} {\bibfnamefont {A.}~\bibnamefont {Mutig}},
  \bibinfo {author} {\bibfnamefont {M.}~\bibnamefont {Kuntz}}, \bibinfo
  {author} {\bibfnamefont {G.}~\bibnamefont {Fiol}}, \bibinfo {author}
  {\bibfnamefont {D.}~\bibnamefont {Bimberg}}, \bibinfo {author} {\bibfnamefont
  {N.~N.}\ \bibnamefont {Ledentsov}}, \bibinfo {author} {\bibfnamefont {V.~A.}\
  \bibnamefont {Shchukin}}, \bibinfo {author} {\bibfnamefont {S.~S.}\
  \bibnamefont {Mikhrin}}, \bibinfo {author} {\bibfnamefont {D.~L.}\
  \bibnamefont {Livshits}}, \bibinfo {author} {\bibfnamefont {I.~L.}\
  \bibnamefont {Krestnikov}}, \bibinfo {author} {\bibfnamefont {A.~R.}\
  \bibnamefont {Kovsh}}, \bibinfo {author} {\bibfnamefont {N.~D.}\ \bibnamefont
  {Zakharov}}, \ and\ \bibinfo {author} {\bibfnamefont {P.}~\bibnamefont
  {Werner}},\ }\href {\doibase 10.1063/1.2358114} {\bibfield  {journal}
  {\bibinfo  {journal} {Applied Physics Letters}\ }\textbf {\bibinfo {volume}
  {89}} (\bibinfo {year} {2006}),\ 10.1063/1.2358114}\BibitemShut {NoStop}%
\bibitem [{\citenamefont {Li}\ \emph {et~al.}(2007)\citenamefont {Li},
  \citenamefont {Patriarche}, \citenamefont {Rossetti},\ and\ \citenamefont
  {Fiore}}]{Li2007GrowthSubstrate}%
  \BibitemOpen
  \bibfield  {author} {\bibinfo {author} {\bibfnamefont {L.~H.}\ \bibnamefont
  {Li}}, \bibinfo {author} {\bibfnamefont {G.}~\bibnamefont {Patriarche}},
  \bibinfo {author} {\bibfnamefont {M.}~\bibnamefont {Rossetti}}, \ and\
  \bibinfo {author} {\bibfnamefont {A.}~\bibnamefont {Fiore}},\ }\href
  {\doibase 10.1063/1.2764212} {\bibfield  {journal} {\bibinfo  {journal}
  {Journal of Applied Physics}\ }\textbf {\bibinfo {volume} {102}},\ \bibinfo
  {pages} {33502} (\bibinfo {year} {2007})}\BibitemShut {NoStop}%
\bibitem [{\citenamefont {He}\ \emph {et~al.}(2004)\citenamefont {He},
  \citenamefont {N{\"{o}}zel}, \citenamefont {Offermans}, \citenamefont
  {Koenraad}, \citenamefont {Gong}, \citenamefont {Hamhuis}, \citenamefont
  {Eijkemans},\ and\ \citenamefont {Wolter}}]{He2004FormationGaAs100}%
  \BibitemOpen
  \bibfield  {author} {\bibinfo {author} {\bibfnamefont {J.}~\bibnamefont
  {He}}, \bibinfo {author} {\bibfnamefont {R.}~\bibnamefont {N{\"{o}}zel}},
  \bibinfo {author} {\bibfnamefont {P.}~\bibnamefont {Offermans}}, \bibinfo
  {author} {\bibfnamefont {P.~M.}\ \bibnamefont {Koenraad}}, \bibinfo {author}
  {\bibfnamefont {Q.}~\bibnamefont {Gong}}, \bibinfo {author} {\bibfnamefont
  {G.~J.}\ \bibnamefont {Hamhuis}}, \bibinfo {author} {\bibfnamefont {T.~J.}\
  \bibnamefont {Eijkemans}}, \ and\ \bibinfo {author} {\bibfnamefont {J.~H.}\
  \bibnamefont {Wolter}},\ }\href {\doibase 10.1063/1.1801172} {\bibfield
  {journal} {\bibinfo  {journal} {Applied Physics Letters}\ }\textbf {\bibinfo
  {volume} {85}},\ \bibinfo {pages} {2771} (\bibinfo {year}
  {2004})}\BibitemShut {NoStop}%
\bibitem [{\citenamefont {Bell}\ \emph {et~al.}(2000)\citenamefont {Bell},
  \citenamefont {Krzyzewski}, \citenamefont {Joyce},\ and\ \citenamefont
  {Jones}}]{Bell2000IslandEffects}%
  \BibitemOpen
  \bibfield  {author} {\bibinfo {author} {\bibfnamefont {G.~R.}\ \bibnamefont
  {Bell}}, \bibinfo {author} {\bibfnamefont {T.~J.}\ \bibnamefont
  {Krzyzewski}}, \bibinfo {author} {\bibfnamefont {P.~B.}\ \bibnamefont
  {Joyce}}, \ and\ \bibinfo {author} {\bibfnamefont {T.~S.}\ \bibnamefont
  {Jones}},\ }\href {\doibase 10.1103/PhysRevB.61.R10551} {\emph {\bibinfo
  {title} {Physical Review B - Condensed Matter and Materials Physics}}},\
  \bibinfo {type} {Tech. Rep.}\ \bibinfo {number} {16}\ (\bibinfo
  {institution} {Centre for Electronic Materials and Devices and Department of
  Chemistry, Imperial College of Science, Technology and Medicine},\ \bibinfo
  {address} {London SW7 2AY, United Kingdom},\ \bibinfo {year}
  {2000})\BibitemShut {NoStop}%
\bibitem [{\citenamefont {Kamiya}\ \emph {et~al.}(1992)\citenamefont {Kamiya},
  \citenamefont {Aspnes}, \citenamefont {Florez},\ and\ \citenamefont
  {Harbison}}]{Kamiya1992Reflectance-differenceVacuum}%
  \BibitemOpen
  \bibfield  {author} {\bibinfo {author} {\bibfnamefont {I.}~\bibnamefont
  {Kamiya}}, \bibinfo {author} {\bibfnamefont {D.~E.}\ \bibnamefont {Aspnes}},
  \bibinfo {author} {\bibfnamefont {L.~T.}\ \bibnamefont {Florez}}, \ and\
  \bibinfo {author} {\bibfnamefont {J.~P.}\ \bibnamefont {Harbison}},\ }\href
  {\doibase 10.1103/PhysRevB.46.15894} {\bibfield  {journal} {\bibinfo
  {journal} {Physical Review B}\ }\textbf {\bibinfo {volume} {46}},\ \bibinfo
  {pages} {15894} (\bibinfo {year} {1992})}\BibitemShut {NoStop}%
\bibitem [{\citenamefont {Belk}\ \emph {et~al.}(1997)\citenamefont {Belk},
  \citenamefont {McConville}, \citenamefont {Sudijono}, \citenamefont {Jones},\
  and\ \citenamefont {Joyce}}]{Belk1997SurfaceEpitaxy}%
  \BibitemOpen
  \bibfield  {author} {\bibinfo {author} {\bibfnamefont {J.~G.}\ \bibnamefont
  {Belk}}, \bibinfo {author} {\bibfnamefont {C.~F.}\ \bibnamefont
  {McConville}}, \bibinfo {author} {\bibfnamefont {J.~L.}\ \bibnamefont
  {Sudijono}}, \bibinfo {author} {\bibfnamefont {T.~S.}\ \bibnamefont {Jones}},
  \ and\ \bibinfo {author} {\bibfnamefont {B.~A.}\ \bibnamefont {Joyce}},\
  }\href {\doibase 10.1016/S0039-6028(97)00355-5} {\bibfield  {journal}
  {\bibinfo  {journal} {Surface Science}\ }\textbf {\bibinfo {volume} {387}},\
  \bibinfo {pages} {213} (\bibinfo {year} {1997})}\BibitemShut {NoStop}%
\bibitem [{\citenamefont {LaBella}\ \emph {et~al.}(2005)\citenamefont
  {LaBella}, \citenamefont {Krause}, \citenamefont {Ding},\ and\ \citenamefont
  {Thibado}}]{LaBella2005Arsenic-richStructure}%
  \BibitemOpen
  \bibfield  {author} {\bibinfo {author} {\bibfnamefont {V.~P.}\ \bibnamefont
  {LaBella}}, \bibinfo {author} {\bibfnamefont {M.~R.}\ \bibnamefont {Krause}},
  \bibinfo {author} {\bibfnamefont {Z.}~\bibnamefont {Ding}}, \ and\ \bibinfo
  {author} {\bibfnamefont {P.~M.}\ \bibnamefont {Thibado}},\ }\href {\doibase
  10.1016/j.surfrep.2005.10.001} {\enquote {\bibinfo {title} {{Arsenic-rich
  GaAs(0 0 1) surface structure}},}\ } (\bibinfo {year} {2005})\BibitemShut
  {NoStop}%
\bibitem [{\citenamefont {Lenz}\ \emph {et~al.}(2010)\citenamefont {Lenz},
  \citenamefont {Eisele}, \citenamefont {Becker}, \citenamefont {Ivanova},
  \citenamefont {Lenz}, \citenamefont {Luckert}, \citenamefont
  {P{\"{o}}tschke}, \citenamefont {Strittmatter}, \citenamefont {Pohl},
  \citenamefont {Bimberg},\ and\ \citenamefont
  {D{\"{a}}hne}}]{Lenz2010AtomicGaAs}%
  \BibitemOpen
  \bibfield  {author} {\bibinfo {author} {\bibfnamefont {A.}~\bibnamefont
  {Lenz}}, \bibinfo {author} {\bibfnamefont {H.}~\bibnamefont {Eisele}},
  \bibinfo {author} {\bibfnamefont {J.}~\bibnamefont {Becker}}, \bibinfo
  {author} {\bibfnamefont {L.}~\bibnamefont {Ivanova}}, \bibinfo {author}
  {\bibfnamefont {E.}~\bibnamefont {Lenz}}, \bibinfo {author} {\bibfnamefont
  {F.}~\bibnamefont {Luckert}}, \bibinfo {author} {\bibfnamefont
  {K.}~\bibnamefont {P{\"{o}}tschke}}, \bibinfo {author} {\bibfnamefont
  {A.}~\bibnamefont {Strittmatter}}, \bibinfo {author} {\bibfnamefont {U.~W.}\
  \bibnamefont {Pohl}}, \bibinfo {author} {\bibfnamefont {D.}~\bibnamefont
  {Bimberg}}, \ and\ \bibinfo {author} {\bibfnamefont {M.}~\bibnamefont
  {D{\"{a}}hne}},\ }\href {\doibase 10.1143/APEX.3.105602} {\bibfield
  {journal} {\bibinfo  {journal} {Applied Physics Express}\ }\textbf {\bibinfo
  {volume} {3}},\ \bibinfo {pages} {105602} (\bibinfo {year}
  {2010})}\BibitemShut {NoStop}%
\bibitem [{\citenamefont {Harrison}\ \emph {et~al.}(2016)\citenamefont
  {Harrison}, \citenamefont {Young}, \citenamefont {Hodgson}, \citenamefont
  {Young}, \citenamefont {Hayne}, \citenamefont {Danos}, \citenamefont
  {Schliwa}, \citenamefont {Strittmatter}, \citenamefont {Lenz}, \citenamefont
  {Eisele}, \citenamefont {Pohl},\ and\ \citenamefont
  {Bimberg}}]{Harrison2016HeterodimensionalGaAs}%
  \BibitemOpen
  \bibfield  {author} {\bibinfo {author} {\bibfnamefont {S.}~\bibnamefont
  {Harrison}}, \bibinfo {author} {\bibfnamefont {M.~P.}\ \bibnamefont {Young}},
  \bibinfo {author} {\bibfnamefont {P.~D.}\ \bibnamefont {Hodgson}}, \bibinfo
  {author} {\bibfnamefont {R.~J.}\ \bibnamefont {Young}}, \bibinfo {author}
  {\bibfnamefont {M.}~\bibnamefont {Hayne}}, \bibinfo {author} {\bibfnamefont
  {L.}~\bibnamefont {Danos}}, \bibinfo {author} {\bibfnamefont
  {A.}~\bibnamefont {Schliwa}}, \bibinfo {author} {\bibfnamefont
  {A.}~\bibnamefont {Strittmatter}}, \bibinfo {author} {\bibfnamefont
  {A.}~\bibnamefont {Lenz}}, \bibinfo {author} {\bibfnamefont {H.}~\bibnamefont
  {Eisele}}, \bibinfo {author} {\bibfnamefont {U.~W.}\ \bibnamefont {Pohl}}, \
  and\ \bibinfo {author} {\bibfnamefont {D.}~\bibnamefont {Bimberg}},\ }\href
  {\doibase 10.1103/PhysRevB.93.085302} {\bibfield  {journal} {\bibinfo
  {journal} {Physical Review B}\ }\textbf {\bibinfo {volume} {93}},\ \bibinfo
  {pages} {1} (\bibinfo {year} {2016})}\BibitemShut {NoStop}%
\bibitem [{\citenamefont {Lingnau}\ \emph {et~al.}(2016)\citenamefont
  {Lingnau}, \citenamefont {L{\"{u}}dge}, \citenamefont {Herzog}, \citenamefont
  {Kolarczik}, \citenamefont {Kaptan}, \citenamefont {Woggon},\ and\
  \citenamefont {Owschimikow}}]{Lingnau2016UltrafastDots}%
  \BibitemOpen
  \bibfield  {author} {\bibinfo {author} {\bibfnamefont {B.}~\bibnamefont
  {Lingnau}}, \bibinfo {author} {\bibfnamefont {K.}~\bibnamefont
  {L{\"{u}}dge}}, \bibinfo {author} {\bibfnamefont {B.}~\bibnamefont {Herzog}},
  \bibinfo {author} {\bibfnamefont {M.}~\bibnamefont {Kolarczik}}, \bibinfo
  {author} {\bibfnamefont {Y.}~\bibnamefont {Kaptan}}, \bibinfo {author}
  {\bibfnamefont {U.}~\bibnamefont {Woggon}}, \ and\ \bibinfo {author}
  {\bibfnamefont {N.}~\bibnamefont {Owschimikow}},\ }\href {\doibase
  10.1103/PhysRevB.94.014305} {\bibfield  {journal} {\bibinfo  {journal}
  {Physical Review B}\ }\textbf {\bibinfo {volume} {94}},\ \bibinfo {pages}
  {14305} (\bibinfo {year} {2016})}\BibitemShut {NoStop}%
\bibitem [{\citenamefont {Han}\ \emph {et~al.}(2019)\citenamefont {Han},
  \citenamefont {Kim}, \citenamefont {Shin}, \citenamefont {Kim}, \citenamefont
  {Noh}, \citenamefont {Lee},\ and\ \citenamefont
  {Krishna}}]{Han2019PhotoluminescenceLayers}%
  \BibitemOpen
  \bibfield  {author} {\bibinfo {author} {\bibfnamefont {I.~S.}\ \bibnamefont
  {Han}}, \bibinfo {author} {\bibfnamefont {J.~S.}\ \bibnamefont {Kim}},
  \bibinfo {author} {\bibfnamefont {J.~C.}\ \bibnamefont {Shin}}, \bibinfo
  {author} {\bibfnamefont {J.~O.}\ \bibnamefont {Kim}}, \bibinfo {author}
  {\bibfnamefont {S.~K.}\ \bibnamefont {Noh}}, \bibinfo {author} {\bibfnamefont
  {S.~J.}\ \bibnamefont {Lee}}, \ and\ \bibinfo {author} {\bibfnamefont
  {S.}~\bibnamefont {Krishna}},\ }\href {\doibase 10.1016/j.jlumin.2018.11.052}
  {\bibfield  {journal} {\bibinfo  {journal} {Journal of Luminescence}\
  }\textbf {\bibinfo {volume} {207}},\ \bibinfo {pages} {512} (\bibinfo {year}
  {2019})}\BibitemShut {NoStop}%
\bibitem [{\citenamefont {Hopfer}\ \emph {et~al.}(2007)\citenamefont {Hopfer},
  \citenamefont {Mutig}, \citenamefont {Fiol}, \citenamefont {Kuntz},
  \citenamefont {Shchukin}, \citenamefont {Haisler}, \citenamefont {Warming},
  \citenamefont {Stock}, \citenamefont {Mikhrin}, \citenamefont {Krestnikov},
  \citenamefont {Livshits}, \citenamefont {Kovsh}, \citenamefont {Bornholdt},
  \citenamefont {Lenz}, \citenamefont {Eisele}, \citenamefont {D{\"{a}}hne},
  \citenamefont {Ledentsov},\ and\ \citenamefont {Bimberg}}]{Hopfer200720Dots}%
  \BibitemOpen
  \bibfield  {author} {\bibinfo {author} {\bibfnamefont {F.}~\bibnamefont
  {Hopfer}}, \bibinfo {author} {\bibfnamefont {A.}~\bibnamefont {Mutig}},
  \bibinfo {author} {\bibfnamefont {G.}~\bibnamefont {Fiol}}, \bibinfo {author}
  {\bibfnamefont {M.}~\bibnamefont {Kuntz}}, \bibinfo {author} {\bibfnamefont
  {V.~A.}\ \bibnamefont {Shchukin}}, \bibinfo {author} {\bibfnamefont {V.~A.}\
  \bibnamefont {Haisler}}, \bibinfo {author} {\bibfnamefont {T.}~\bibnamefont
  {Warming}}, \bibinfo {author} {\bibfnamefont {E.}~\bibnamefont {Stock}},
  \bibinfo {author} {\bibfnamefont {S.~S.}\ \bibnamefont {Mikhrin}}, \bibinfo
  {author} {\bibfnamefont {I.~L.}\ \bibnamefont {Krestnikov}}, \bibinfo
  {author} {\bibfnamefont {D.~A.}\ \bibnamefont {Livshits}}, \bibinfo {author}
  {\bibfnamefont {A.~R.}\ \bibnamefont {Kovsh}}, \bibinfo {author}
  {\bibfnamefont {C.}~\bibnamefont {Bornholdt}}, \bibinfo {author}
  {\bibfnamefont {A.}~\bibnamefont {Lenz}}, \bibinfo {author} {\bibfnamefont
  {H.}~\bibnamefont {Eisele}}, \bibinfo {author} {\bibfnamefont
  {M.}~\bibnamefont {D{\"{a}}hne}}, \bibinfo {author} {\bibfnamefont {N.~N.}\
  \bibnamefont {Ledentsov}}, \ and\ \bibinfo {author} {\bibfnamefont
  {D.}~\bibnamefont {Bimberg}},\ }in\ \href {\doibase
  10.1109/JSTQE.2007.905133} {\emph {\bibinfo {booktitle} {IEEE Journal on
  Selected Topics in Quantum Electronics}}},\ Vol.~\bibinfo {volume} {13}\
  (\bibinfo {year} {2007})\ pp.\ \bibinfo {pages} {1302--1308}\BibitemShut
  {NoStop}%
\bibitem [{\citenamefont {Ledentsov}\ \emph {et~al.}(2007)\citenamefont
  {Ledentsov}, \citenamefont {Bimberg}, \citenamefont {Hopfer}, \citenamefont
  {Mutig}, \citenamefont {Shchukin}, \citenamefont {Savel'Ev}, \citenamefont
  {Fiol}, \citenamefont {Stock}, \citenamefont {Eisele}, \citenamefont
  {D{\"{a}}hne}, \citenamefont {Gerthsen}, \citenamefont {Fischer},
  \citenamefont {Litvinov}, \citenamefont {Rosenauer}, \citenamefont {Mikhrin},
  \citenamefont {Kovsh}, \citenamefont {Zakharov},\ and\ \citenamefont
  {Werner}}]{Ledentsov2007SubmonolayerLasers}%
  \BibitemOpen
  \bibfield  {author} {\bibinfo {author} {\bibfnamefont {N.~N.}\ \bibnamefont
  {Ledentsov}}, \bibinfo {author} {\bibfnamefont {D.}~\bibnamefont {Bimberg}},
  \bibinfo {author} {\bibfnamefont {F.}~\bibnamefont {Hopfer}}, \bibinfo
  {author} {\bibfnamefont {A.}~\bibnamefont {Mutig}}, \bibinfo {author}
  {\bibfnamefont {V.~A.}\ \bibnamefont {Shchukin}}, \bibinfo {author}
  {\bibfnamefont {A.~V.}\ \bibnamefont {Savel'Ev}}, \bibinfo {author}
  {\bibfnamefont {G.}~\bibnamefont {Fiol}}, \bibinfo {author} {\bibfnamefont
  {E.}~\bibnamefont {Stock}}, \bibinfo {author} {\bibfnamefont
  {H.}~\bibnamefont {Eisele}}, \bibinfo {author} {\bibfnamefont
  {M.}~\bibnamefont {D{\"{a}}hne}}, \bibinfo {author} {\bibfnamefont
  {D.}~\bibnamefont {Gerthsen}}, \bibinfo {author} {\bibfnamefont
  {U.}~\bibnamefont {Fischer}}, \bibinfo {author} {\bibfnamefont
  {D.}~\bibnamefont {Litvinov}}, \bibinfo {author} {\bibfnamefont
  {A.}~\bibnamefont {Rosenauer}}, \bibinfo {author} {\bibfnamefont {S.~S.}\
  \bibnamefont {Mikhrin}}, \bibinfo {author} {\bibfnamefont {A.~R.}\
  \bibnamefont {Kovsh}}, \bibinfo {author} {\bibfnamefont {N.~D.}\ \bibnamefont
  {Zakharov}}, \ and\ \bibinfo {author} {\bibfnamefont {P.}~\bibnamefont
  {Werner}},\ }\href {\doibase 10.1007/s11671-007-9078-0} {\bibfield  {journal}
  {\bibinfo  {journal} {Nanoscale Research Letters}\ }\textbf {\bibinfo
  {volume} {2}},\ \bibinfo {pages} {417} (\bibinfo {year} {2007})}\BibitemShut
  {NoStop}%
\bibitem [{\citenamefont {Mikhrin}\ \emph {et~al.}(2000)\citenamefont
  {Mikhrin}, \citenamefont {Zhukov}, \citenamefont {Kovsh}, \citenamefont
  {Maleev}, \citenamefont {Ustinov}, \citenamefont {Shernyakov}, \citenamefont
  {Soshnikov}, \citenamefont {Livshits}, \citenamefont {Tarasov}, \citenamefont
  {Bedarev}, \citenamefont {Volovik}, \citenamefont {Maximov}, \citenamefont
  {Tsatsul'nikov}, \citenamefont {Ledentsov}, \citenamefont {Kop'ev},
  \citenamefont {Bimberg},\ and\ \citenamefont
  {Alferov}}]{Mikhrin20000.94Dots}%
  \BibitemOpen
  \bibfield  {author} {\bibinfo {author} {\bibfnamefont {S.~S.}\ \bibnamefont
  {Mikhrin}}, \bibinfo {author} {\bibfnamefont {A.~E.}\ \bibnamefont {Zhukov}},
  \bibinfo {author} {\bibfnamefont {A.~R.}\ \bibnamefont {Kovsh}}, \bibinfo
  {author} {\bibfnamefont {N.~A.}\ \bibnamefont {Maleev}}, \bibinfo {author}
  {\bibfnamefont {V.~M.}\ \bibnamefont {Ustinov}}, \bibinfo {author}
  {\bibfnamefont {Y.~M.}\ \bibnamefont {Shernyakov}}, \bibinfo {author}
  {\bibfnamefont {I.~P.}\ \bibnamefont {Soshnikov}}, \bibinfo {author}
  {\bibfnamefont {D.~A.}\ \bibnamefont {Livshits}}, \bibinfo {author}
  {\bibfnamefont {I.~S.}\ \bibnamefont {Tarasov}}, \bibinfo {author}
  {\bibfnamefont {D.~A.}\ \bibnamefont {Bedarev}}, \bibinfo {author}
  {\bibfnamefont {B.~V.}\ \bibnamefont {Volovik}}, \bibinfo {author}
  {\bibfnamefont {M.~V.}\ \bibnamefont {Maximov}}, \bibinfo {author}
  {\bibfnamefont {A.~F.}\ \bibnamefont {Tsatsul'nikov}}, \bibinfo {author}
  {\bibfnamefont {N.~N.}\ \bibnamefont {Ledentsov}}, \bibinfo {author}
  {\bibfnamefont {P.~S.}\ \bibnamefont {Kop'ev}}, \bibinfo {author}
  {\bibfnamefont {D.}~\bibnamefont {Bimberg}}, \ and\ \bibinfo {author}
  {\bibfnamefont {Z.~I.}\ \bibnamefont {Alferov}},\ }\href {\doibase
  10.1088/0268-1242/15/11/309} {\bibfield  {journal} {\bibinfo  {journal}
  {Semiconductor Science and Technology}\ }\textbf {\bibinfo {volume} {15}},\
  \bibinfo {pages} {1061} (\bibinfo {year} {2000})}\BibitemShut {NoStop}%
\bibitem [{\citenamefont {Alzeidan}\ \emph
  {et~al.}(2019{\natexlab{a}})\citenamefont {Alzeidan}, \citenamefont {Claro},\
  and\ \citenamefont {Quivy}}]{Alzeidan2019High-detectivityReconstruction}%
  \BibitemOpen
  \bibfield  {author} {\bibinfo {author} {\bibfnamefont {A.}~\bibnamefont
  {Alzeidan}}, \bibinfo {author} {\bibfnamefont {M.~S.}\ \bibnamefont {Claro}},
  \ and\ \bibinfo {author} {\bibfnamefont {A.~A.}\ \bibnamefont {Quivy}},\
  }\href {\doibase 10.1063/1.5125238} {\bibfield  {journal} {\bibinfo
  {journal} {Journal of Applied Physics}\ }\textbf {\bibinfo {volume} {126}},\
  \bibinfo {pages} {224506} (\bibinfo {year} {2019}{\natexlab{a}})}\BibitemShut
  {NoStop}%
\bibitem [{\citenamefont {Alzeidan}\ \emph
  {et~al.}(2019{\natexlab{b}})\citenamefont {Alzeidan}, \citenamefont
  {Cantalice}, \citenamefont {Garcia}, \citenamefont {Deneke},\ and\
  \citenamefont {Quivy}}]{Alzeidan2019InvestigationPhotodetector}%
  \BibitemOpen
  \bibfield  {author} {\bibinfo {author} {\bibfnamefont {A.}~\bibnamefont
  {Alzeidan}}, \bibinfo {author} {\bibfnamefont {T.~F.}\ \bibnamefont
  {Cantalice}}, \bibinfo {author} {\bibfnamefont {A.~J.}\ \bibnamefont
  {Garcia}}, \bibinfo {author} {\bibfnamefont {C.~F.}\ \bibnamefont {Deneke}},
  \ and\ \bibinfo {author} {\bibfnamefont {A.~A.}\ \bibnamefont {Quivy}},\ }in\
  \href {\doibase 10.1109/SBMicro.2019.8919349} {\emph {\bibinfo {booktitle}
  {SBMicro 2019 - 34th Symposium on Microelectronics Technology and Devices}}}\
  (\bibinfo  {publisher} {Institute of Electrical and Electronics Engineers
  Inc.},\ \bibinfo {year} {2019})\BibitemShut {NoStop}%
\bibitem [{\citenamefont {Lam}\ \emph {et~al.}(2014)\citenamefont {Lam},
  \citenamefont {Wu}, \citenamefont {Tang}, \citenamefont {Jiang},
  \citenamefont {Hatch}, \citenamefont {Beanland}, \citenamefont {Wilson},
  \citenamefont {Allison},\ and\ \citenamefont
  {Liu}}]{Lam2014SubmonolayerCells}%
  \BibitemOpen
  \bibfield  {author} {\bibinfo {author} {\bibfnamefont {P.}~\bibnamefont
  {Lam}}, \bibinfo {author} {\bibfnamefont {J.}~\bibnamefont {Wu}}, \bibinfo
  {author} {\bibfnamefont {M.}~\bibnamefont {Tang}}, \bibinfo {author}
  {\bibfnamefont {Q.}~\bibnamefont {Jiang}}, \bibinfo {author} {\bibfnamefont
  {S.}~\bibnamefont {Hatch}}, \bibinfo {author} {\bibfnamefont
  {R.}~\bibnamefont {Beanland}}, \bibinfo {author} {\bibfnamefont
  {J.}~\bibnamefont {Wilson}}, \bibinfo {author} {\bibfnamefont
  {R.}~\bibnamefont {Allison}}, \ and\ \bibinfo {author} {\bibfnamefont
  {H.}~\bibnamefont {Liu}},\ }\href {\doibase 10.1016/j.solmat.2014.03.046}
  {\bibfield  {journal} {\bibinfo  {journal} {Solar Energy Materials and Solar
  Cells}\ }\textbf {\bibinfo {volume} {126}},\ \bibinfo {pages} {83} (\bibinfo
  {year} {2014})}\BibitemShut {NoStop}%
\bibitem [{\citenamefont {Feenstra}\ \emph {et~al.}(1987)\citenamefont
  {Feenstra}, \citenamefont {Stroscio}, \citenamefont {Tersoff},\ and\
  \citenamefont {Fein}}]{PhysRevLett.58.1192}%
  \BibitemOpen
  \bibfield  {author} {\bibinfo {author} {\bibfnamefont {R.~M.}\ \bibnamefont
  {Feenstra}}, \bibinfo {author} {\bibfnamefont {J.~A.}\ \bibnamefont
  {Stroscio}}, \bibinfo {author} {\bibfnamefont {J.}~\bibnamefont {Tersoff}}, \
  and\ \bibinfo {author} {\bibfnamefont {A.~P.}\ \bibnamefont {Fein}},\ }\href
  {\doibase 10.1103/PhysRevLett.58.1192} {\bibfield  {journal} {\bibinfo
  {journal} {Physical Review Letters}\ }\textbf {\bibinfo {volume} {58}},\
  \bibinfo {pages} {1192} (\bibinfo {year} {1987})}\BibitemShut {NoStop}%
\bibitem [{\citenamefont {Kim}\ \emph {et~al.}(2015)\citenamefont {Kim},
  \citenamefont {Ban},\ and\ \citenamefont
  {Honsberg}}]{Kim2015Multi-stackedCells}%
  \BibitemOpen
  \bibfield  {author} {\bibinfo {author} {\bibfnamefont {Y.}~\bibnamefont
  {Kim}}, \bibinfo {author} {\bibfnamefont {K.~Y.}\ \bibnamefont {Ban}}, \ and\
  \bibinfo {author} {\bibfnamefont {C.~B.}\ \bibnamefont {Honsberg}},\ }\href
  {\doibase 10.1063/1.4922274} {\bibfield  {journal} {\bibinfo  {journal}
  {Applied Physics Letters}\ }\textbf {\bibinfo {volume} {106}},\ \bibinfo
  {pages} {222104} (\bibinfo {year} {2015})}\BibitemShut {NoStop}%
\bibitem [{\citenamefont {Muraki}\ \emph {et~al.}(1992)\citenamefont {Muraki},
  \citenamefont {Fukatsu}, \citenamefont {Shiraki},\ and\ \citenamefont
  {Ito}}]{Muraki1992SurfaceWells}%
  \BibitemOpen
  \bibfield  {author} {\bibinfo {author} {\bibfnamefont {K.}~\bibnamefont
  {Muraki}}, \bibinfo {author} {\bibfnamefont {S.}~\bibnamefont {Fukatsu}},
  \bibinfo {author} {\bibfnamefont {Y.}~\bibnamefont {Shiraki}}, \ and\
  \bibinfo {author} {\bibfnamefont {R.}~\bibnamefont {Ito}},\ }\href {\doibase
  10.1063/1.107835} {\bibfield  {journal} {\bibinfo  {journal} {Applied Physics
  Letters}\ }\textbf {\bibinfo {volume} {61}},\ \bibinfo {pages} {557}
  (\bibinfo {year} {1992})}\BibitemShut {NoStop}%
\bibitem [{\citenamefont {Martini}\ \emph {et~al.}(2003)\citenamefont
  {Martini}, \citenamefont {Quivy}, \citenamefont {Lamas}, \citenamefont
  {Da~Silva}, \citenamefont {Da~Silva},\ and\ \citenamefont
  {Leite}}]{Martini2003InfluenceSurface}%
  \BibitemOpen
  \bibfield  {author} {\bibinfo {author} {\bibfnamefont {S.}~\bibnamefont
  {Martini}}, \bibinfo {author} {\bibfnamefont {A.~A.}\ \bibnamefont {Quivy}},
  \bibinfo {author} {\bibfnamefont {T.~E.}\ \bibnamefont {Lamas}}, \bibinfo
  {author} {\bibfnamefont {M.~J.}\ \bibnamefont {Da~Silva}}, \bibinfo {author}
  {\bibfnamefont {E.~C.}\ \bibnamefont {Da~Silva}}, \ and\ \bibinfo {author}
  {\bibfnamefont {J.~R.}\ \bibnamefont {Leite}},\ }in\ \href {\doibase
  10.1016/S0022-0248(02)02313-8} {\emph {\bibinfo {booktitle} {Journal of
  Crystal Growth}}},\ Vol.\ \bibinfo {volume} {251}\ (\bibinfo {year} {2003})\
  pp.\ \bibinfo {pages} {101--105}\BibitemShut {NoStop}%
\bibitem [{\citenamefont {Cantalice}\ \emph {et~al.}(2019)\citenamefont
  {Cantalice}, \citenamefont {Alzeidan}, \citenamefont {Urahata},\ and\
  \citenamefont {Quivy}}]{Cantalice2019In-situDots}%
  \BibitemOpen
  \bibfield  {author} {\bibinfo {author} {\bibfnamefont {T.~F.}\ \bibnamefont
  {Cantalice}}, \bibinfo {author} {\bibfnamefont {A.}~\bibnamefont {Alzeidan}},
  \bibinfo {author} {\bibfnamefont {S.~M.}\ \bibnamefont {Urahata}}, \ and\
  \bibinfo {author} {\bibfnamefont {A.~A.}\ \bibnamefont {Quivy}},\ }\href
  {\doibase 10.1088/2053-1591/ab55a8} {\bibfield  {journal} {\bibinfo
  {journal} {Mater. Res. Express}\ }\textbf {\bibinfo {volume} {6}},\ \bibinfo
  {pages} {126205} (\bibinfo {year} {2019})}\BibitemShut {NoStop}%
\bibitem [{\citenamefont {U{\v{z}}davinys}\ \emph {et~al.}(2017)\citenamefont
  {U{\v{z}}davinys}, \citenamefont {Becerra}, \citenamefont {Ivanov},
  \citenamefont {DenBaars}, \citenamefont {Nakamura}, \citenamefont {Speck},\
  and\ \citenamefont
  {Marcinkevi{\v{c}}ius}}]{Uzdavinys2017InfluencePhotoluminescence}%
  \BibitemOpen
  \bibfield  {author} {\bibinfo {author} {\bibfnamefont {T.~K.}\ \bibnamefont
  {U{\v{z}}davinys}}, \bibinfo {author} {\bibfnamefont {D.~L.}\ \bibnamefont
  {Becerra}}, \bibinfo {author} {\bibfnamefont {R.}~\bibnamefont {Ivanov}},
  \bibinfo {author} {\bibfnamefont {S.~P.}\ \bibnamefont {DenBaars}}, \bibinfo
  {author} {\bibfnamefont {S.}~\bibnamefont {Nakamura}}, \bibinfo {author}
  {\bibfnamefont {J.~S.}\ \bibnamefont {Speck}}, \ and\ \bibinfo {author}
  {\bibfnamefont {S.}~\bibnamefont {Marcinkevi{\v{c}}ius}},\ }\href {\doibase
  10.1364/ome.7.003116} {\bibfield  {journal} {\bibinfo  {journal} {Optical
  Materials Express}\ }\textbf {\bibinfo {volume} {7}},\ \bibinfo {pages}
  {3116} (\bibinfo {year} {2017})}\BibitemShut {NoStop}%
\bibitem [{\citenamefont {Pohl}\ \emph {et~al.}(2005)\citenamefont {Pohl},
  \citenamefont {P{\"{o}}tschke}, \citenamefont {Schliwa}, \citenamefont
  {Guffarth}, \citenamefont {Bimberg}, \citenamefont {Zakharov}, \citenamefont
  {Werner}, \citenamefont {Lifshits}, \citenamefont {Shchukin},\ and\
  \citenamefont {Jesson}}]{Pohl2005EvolutionDots}%
  \BibitemOpen
  \bibfield  {author} {\bibinfo {author} {\bibfnamefont {U.~W.}\ \bibnamefont
  {Pohl}}, \bibinfo {author} {\bibfnamefont {K.}~\bibnamefont
  {P{\"{o}}tschke}}, \bibinfo {author} {\bibfnamefont {A.}~\bibnamefont
  {Schliwa}}, \bibinfo {author} {\bibfnamefont {F.}~\bibnamefont {Guffarth}},
  \bibinfo {author} {\bibfnamefont {D.}~\bibnamefont {Bimberg}}, \bibinfo
  {author} {\bibfnamefont {N.~D.}\ \bibnamefont {Zakharov}}, \bibinfo {author}
  {\bibfnamefont {P.}~\bibnamefont {Werner}}, \bibinfo {author} {\bibfnamefont
  {M.~B.}\ \bibnamefont {Lifshits}}, \bibinfo {author} {\bibfnamefont {V.~A.}\
  \bibnamefont {Shchukin}}, \ and\ \bibinfo {author} {\bibfnamefont {D.~E.}\
  \bibnamefont {Jesson}},\ }\href {\doibase 10.1103/PhysRevB.72.245332}
  {\bibfield  {journal} {\bibinfo  {journal} {Physical Review B - Condensed
  Matter and Materials Physics}\ }\textbf {\bibinfo {volume} {72}} (\bibinfo
  {year} {2005}),\ 10.1103/PhysRevB.72.245332}\BibitemShut {NoStop}%
\end{thebibliography}%

\end{document}